\documentclass[12pt]{article}
\usepackage{amsfonts}
\usepackage{mathrsfs}

\usepackage{multirow}

\usepackage{bbding}
\usepackage{amssymb}
\usepackage{amsmath}
\usepackage{graphicx,color}

\newcommand{\bea}{\begin{eqnarray}}
\newcommand{\eea}{\end{eqnarray}}
\newcommand{\Bea}{\begin{eqnarray*}}
\newcommand{\Eea}{\end{eqnarray*}}
\newcommand{\ba}{\begin{array}}
\newcommand{\ea}{\end{array}}
\newcommand{\bt}{\begin{tabular}}
\newcommand{\et}{\end{tabular}}
\newcommand{\btb}{\begin{table}}
\newcommand{\etb}{\end{table}}
\newcommand{\bc}{\begin{center}}
\newcommand{\ec}{\end{center}}
\newcommand{\beq}{\begin{equation}}
\newcommand{\eeq}{\end{equation}}

\begin{document}

\title{ Adaptive post-Dantzig estimation and prediction for non-sparse ``large $p$ and small $n$"
models}
\author{
Lu Lin,  Lixing Zhu\footnote{Lu Lin is a professor of the School of Mathematics
at Shandong University, Jinan, China. His research was supported by
NNSF project (10771123) of China, NBRP (973 Program 2007CB814901) of
China,  RFDP (20070422034) of China, NSF projects (Y2006A13 and
Q2007A05) of Shandong Province of China. Lixing Zhu is a chair
professor of Department of Mathematics at Hong Kong Baptist
University, Hong Kong, China. Email: lzhu@hkbu.edu.hk. He was supported by a grant from the
University Grants Council of Hong Kong, Hong Kong, China. Yujie Gai
is a PHD student of the School of Mathematics at Shandong
University, Jinan, China. The first two authors are in charge of the
methodology development and material organization.}\, \,
  and Yujie Gai}
\date{}
\maketitle

\vspace{-3ex}

\begin{abstract}

For consistency (even oracle properties) of estimation and model prediction, almost all existing methods of variable/feature selection critically depend on sparsity of models. However, for ``large $p$ and small $n$"  models sparsity  assumption is hard to check and particularly, when this assumption is violated, the consistency of all existing estimations  is usually impossible because  working models selected by existing methods such as the LASSO and the Dantzig selector are usually biased.
To attack this problem, we in this paper propose adaptive post-Dantzig estimation and model
prediction. Here the
adaptability means that the consistency based on the newly proposed method is adaptive to
non-sparsity of model, choice of shrinkage
tuning parameter and   dimension of  predictor vector.
The idea is that after a sub-model as a working model is determined by the Dantzig selector, we construct a globally
unbiased sub-model by choosing suitable instrumental variables and
nonparametric adjustment. The new estimation of the parameters in
the sub-model can be of the  asymptotic normality. The
consistent estimator, together with the selected sub-model
 and adjusted model, improves model predictions. Simulation studies show that the new approach has the significant
improvement of estimation and prediction accuracies over the Gaussian Dantzig selector
and other classical methods have.

\

{\it Keywords.} Adaptability, bias correction, Dantzig selector, instrumental variable,  nonparametric
adjustment, Ultra high-dimensional regression.

\vspace{1ex}

{\it AMS 2001 subject classification}: 62C05, 62F10, 62F12, 62G05.

\vspace{1ex}

{\it Running head.} Adaptive post-Dantzig inference.

\end{abstract}

\baselineskip=21pt

\newpage

\noindent {\large\bf 1. Introduction}

Estimation consistency is a natural criterion for estimation
accuracy. In classical settings with small/moderate number of
variables in models, this criterion can be adopted. For high-dimensional models, particularly, when the number
$p$ of variables involved is even larger than the sample size $n$, are called
 ``large $p$, small $n$" models.  However in these paradigm estimation  consistency  becomes a very challenging issue. This
is because what we can work on is only  working models rather than
full models after active variables are selected into
working models. For variable selection, some classical and newly
proposed methods are available, such as the LASSO (including the
adaptive LASSO) and the SCAD. These methods provide consistent and
asymptotically normally distributed estimation for the parameters in
working models, but these properties heavily depend on  sparse
structure,  proper choice of shrinkage tuning parameter and the
diverging rate of the dimension of parameter vector. For the
relevant references see Huber (1973), Portnoy (1988), Bai and
Saranadasa (1996), Fan and Peng (2004), Fan, Peng and Huang (2005),
Lam and Fan (2008), Huang {\it et al.} (2008), and Li, Zhu and Lin
(2009), among others. As  such, for models without spare structure,
how to construct consistent estimation is a great challenge. It is
required to develop new or extended statistical methodologies and
theories to handle this challenge; see for example Donoho (2000),
Kettenring, Lindsay and Siegmund (2003).

To this end, we further review existing methods to get motivation
for new methodology development. The following methods were
developed also under sparse structure. The Dantzig selector that was
proposed by Cand\'es and Tao (2007) and was extended to handle the
generalized linear models by James and Radchenko (2009) has received
much attention. The connection between the Dantzig selector and the
LASSO was investigated by James {\it et al.} (2009).
Under  the uniform uncertainty principle, the resulting estimator
achieves an ideal risk of order $O(\sigma\sqrt{\log p})$ with a
large probability. This implies that for large $p$, such a risk can
be however large and then even under sparse structure the estimator
may also be inconsistent. To reduce the risk and improve the
performance of relevant estimation, the Gaussian Dantzig Selector, a
two-stage estimation, was suggested in the literature (Cand\'es and
Tao 2007). Such an improved estimation is still inconsistent when
the shrinkage tuning parameter is chosen to be large (for details
see the next section). Another method is the Double Dantzig Selector
(James and Radchenko 2009), by which one may choose a more accurate
model and, at the same time,  get a more  accurate estimator. But it
critically depends on the choice of shrinkage tuning parameter.
Motivated by these problems, Fan and Lv (2008) introduced a sure
independent screening method that is based on correlation learning
to reduce high dimensionality   to a moderate scale below the sample
size. Afterwards,  variable selection and parameter estimation can
be accomplished by sophisticated methods, such as the LASSO, the
SCAD or the Dantzig selector. The relevant references include
Kosorok and Ma (2007), Van Der Lanin and Bryan (2001), Chen and Qin
(2010), James, Radchenko and Lv (2009) and Kuelbs and Anand (2010),
among others.

However, for  any model with very large $p$, without model sparsity, all existing methods cannot provide estimation consistency
 for working models, and any further data analysis would be questionable
  unless we can correct biases later or at most we can obtain an approximation
  rather than estimation consistency as the sample size goes to infinity.
To deal with this problem,  we focus our attention on   working
sub-model that is chosen by  the Dantzig selector.  In this paper,
we suggest a method to construct consistent and asymptotically
normal distributed estimation for the parameters in the sub-model.
To achieve this, a nonparametric adjustment is recommended to
construct a globally unbiased sub-model and to correct the bias in
working model. Here the nonparametric adjustment may depend on a
low-dimensional nonparametric estimation via using  proper
instrument variables.  We will show the following properties. The
estimator $\hat\theta$ of the parameter vector $\theta$ in the
sub-model satisfies $\|\hat\theta-\theta\|^2_{{\ell}_2}=O_p(n^{-1})$
and the asymptotic normality if the dimension $q$ of $\theta$ is
fixed. Even for the case where $q$ tends to infinity, the consistent
and asymptotic normality still hold when  $q$ diverges at a certain rate. We will briefly discuss the theoretical
results for the case with diverging $q$. Furthermore, the new
consistent estimator, together with the unbiased adjustment
sub-model or the original sub-model, can also improve model
prediction accuracy. We will prove that  our method possesses the
adaptability. That is, the above properties always hold whether the
sub-model is small or large, the dimension of the parameter in the
original model is high or not, and the original model is sparse or
not.

The rest of the paper is organized as follows. In Section~2  the
properties of the Dantzig estimator for the high-dimensional linear
model are re-examined. In Section~3 a bias-corrected sub-model is
proposed via introducing instrumental variables and a nonparametric
adjustment, and a method about  instrumental variable selection is
introduced. Estimation and prediction procedures for the new
sub-model are suggested and the asymptotic properties of the
resulting estimator and prediction are obtained. In Section~4 the
algorithms for constructing instrumental variables are proposed.
Simulation studies are presented in  Section~5 to examine the
performance of the new approach when compared with the classical
Dantzig selector and other methods. The technical proofs for the
theoretical results are postponed to the Appendix.

\

\

\

\noindent {\large\bf 2. A brief review for the Dantzig selector}

Consider the  model
$$Y=\beta'X+\varepsilon,\eqno(2.1)$$ where $Y$ is the scale
response, $X$ is the $p$-dimensional covariate and $\varepsilon$ is
the random error satisfying $E(\varepsilon|X)=0$ and
$Cov(\varepsilon|X)=\sigma^2$. Here $p$ will be greater than  $n$ when we can collect a sample of size $n$.
Throughout this paper, our primary interest is to construct
consistent estimators for  significant components of the
parameter vector
$\beta=(\beta_1,\cdots,\beta_p)'\in\mathscr{B}\subset R^p$. These
significant components of $\beta$, together with the corresponding
covariates, composes a working model. Then the second interest of
our paper is to obtain reasonable model prediction via our estimation.

To introduce the new estimation, we first re-examine the Dantzig
selector. Let ${\bf Y}=(Y_1,\cdots,Y_n)'$ be the vector
of the observed responses and ${\bf X}=(X_1,\cdots,X_n)'=({\bf x}_1,\cdots,{\bf x}_p)$
be the
 $n\times p$ matrix of the observed covariates. The Dantzig selector of $\beta$ is defined as
$$\tilde \beta^D=\arg\min_{\beta\in\mathscr{B}}\|\beta\|_{{\ell}_1} \ \
\mbox{ subject to } \ \ \sup_{1\leq j\leq p}|{\bf x}'_jr| \leq
\lambda_p\,\sigma\eqno(2.2)$$ for some $\lambda_p>0$, where
$\|\beta\|_{{\ell}_1}=\sum_{j=1}^p|\beta_j|$ and $r={\bf Y}-{\bf
X}\beta$. As was shown by Cand\'es and Tao (2007), under some regularity
conditions, this estimator satisfies that, with large probability,
$$\|\tilde\beta^D-\beta\|^2_{{\ell}_2}\leq  C  \sigma^2\log p,\eqno(2.3)$$
where $C$ is free of $p$ and $\|\tilde\beta^D-\beta\|^2_{{\ell}_2}=
\sum_{j=1}^p|\tilde \beta_j^D-\beta_j|^2$. In fact this is an ideal
risk and thus cannot be improved in a certain sense. However,  such
a risk can become large and may not be negligible when the dimension
$p>n$.

To reduce the risk and enhance the performance in practical
settings, one often uses a two-stage selection procedure (e. g., the Gaussian
Dantzig Selector) to construct a risk-reduced estimator for the
obtained sub-model (Cand\'es and Tao 2007). For example, we can
first estimate $I=\{j:\beta_j\neq 0\}$ with $\tilde I=\{j:|\tilde
\beta_j^D|>\varsigma\sigma\}$ for some $\varsigma\geq 0$ and then
construct an estimator
$$\tilde\beta_{(\tilde I)}=(({\bf X}^{(\tilde I)})'{\bf X}^{(\tilde I)})^{-1}({\bf X}^{(\tilde
I)})'{\bf Y}$$ for $\beta_{(\tilde I)}$ and set the other components
of $\beta$ to be zero, where $\beta_{(\tilde I)}$ is the restriction
of $\beta$ to the set ${\tilde I}$, and ${\bf X}^{(\tilde I)}$ is
the matrix with the column vectors according to $\tilde I$.

Denote $\beta_{(\tilde I)}=\theta$, a $q$-dimensional vector of
interest. Without loss of generality, suppose that $\beta$ can be
partitioned as $\beta=(\theta',\gamma')'$ and, correspondingly, $X$
is partitioned as $X=(Z',U')'$. Then the above two-stage procedure
implies that we can use the sub-model
$$Y=\theta'Z+\eta\eqno(2.4)$$ to replace the full-model (2.1), where
$\eta=\gamma'U+\varepsilon$ is regarded as error. Here the dimension
$q$ of $\theta$ can be either fixed or diverging with $n$ at certain rate. Since
the above sub-model is a replacer of the full model (2.1), we call
$\theta$ and $Z$ the main parts of $\beta$ and $X$, respectively.
From (2.1) and (2.4) it follows that $E(\eta|Z)= \gamma'E(U|Z)$.
When both $\gamma \neq 0$ and $E(U|Z)\neq 0$, the sub-model (2.4) is
biased and thus the two-stage estimator
$\tilde\theta_S=\tilde\beta_{(\tilde I)}$ is also biased. It shows
that the two-stage estimator $\tilde\theta_S$ of
$\theta$ is also inconsistent. Note that for any non-sparse model,
the condition $\gamma \neq 0$ always holds. Then the above method is not possible to  obtain consistent estimation.

Another method for improving the Dantzig selector is the Double
Dantzig Selector. By which more accurate model and estimation can be expected. In the first step, the
Dantzig selector is used with a relatively large shrinkage tuning
parameter $\lambda_p$ defined above to get a relatively accurate
sub-model in the sense that more significant variables are
contained. The Dantzig selector is further used in the selected sub-model to obtain a relatively accurate estimator of $\theta$ via a
small $\lambda_p$ and data $(Y,Z)$.  However, such a method cannot handle non-sparse model because the sub-model selected in the first step has already been biased. It is also
noted that this method critically depends on twice choices of
shrinkage tuning parameter $\lambda_p$; for details see James and
Radchenko (2009). On the other hand, when estimation consistency
and normality, rather than variable selection,  heavily
depend on the choice of $\lambda_p$,
it is practically not convenient, and more seriously, the consistency is in effect not judgeable unless
a criterion of tuning parameter selection can be defined
to ensure consistency.  %
   Then it is desirable to have a new estimation/inference
method with which consistency is free of  the choice of $\lambda_p$.

\noindent {\large\bf 3. Adaptive post-Dantzig estimation and prediction}

\noindent{\bf 3.1 Bias-corrected model.} As was shown above, the
sub-model (2.4) is usually biased. Furthermore, this model is regarded as a
non-random model after the variable selection given by the Dantzig
selector, i.e., the estimate $\tilde I$ for the index set $I$
defined in the previous section is fixed after variable
selection.

It is clear that a bias correction is needed for the selected
sub-model (2.4) when we want to have a consistent estimation of the
sub-vector $\theta=(\theta_1,\cdots,\theta_q)'$.  To this end,  a
new model with an instrumental variable is established. Denote
$Z^\star=(Z',U^{(1)},\cdots,U^{(d)})'$ and $W=AZ^\star$, where $A$
is $d\times (q+d)$ matrix satisfying that its row vectors have
length 1. Without loss of generality, $U^{(1)},\cdots,U^{(d)}$ are
supposed to be the first $d$ components of $U$, although they may be
chosen as another components of $U$ or pseudo-variables (artificial
vavriables). Denote by $\lambda_M$ the maximum eigenvalue of $UU'$
and set $V=(\alpha'U/\rho,W')'$ for some $\alpha$ to be chosen
later, where $\rho$ is a nonrandom positive number satisfying the
condition
 $\rho=O(\|\alpha\|_{{\ell}_2}\sqrt{\lambda_M})$. Choose
$A$ and $U^{(1)},\cdots,U^{(d)}$ such that
$$E\{(Z-E(Z|V))(Z-E(Z|V))'\}>0.\eqno(3.1)$$
This condition on the matrix we need can trivially hold because $V$
contains $W$ that is a weighted sum of $Z$ and
$U^{(1)},\cdots,U^{(d)}$. The condition (3.1) can be used to guarantee
the identifiability of the following model.

Denote $g(V)=E(\eta|V)$.  Now we introduce a
bias-corrected version of (2.4) as
$$Y_i=\theta'Z_i+g(V_i)+\xi(V_i),\ i=1,\cdots,n, \eqno(3.2)$$ where $\xi(V)
=\eta-g(V)$. Obviously, if $\alpha$ in $V$ is identical to $\gamma$
in $\eta$, this model is unbiased, i.e., $E(\xi|Z,V)=0$; otherwise
it may be biased. This model can be regarded as a partially linear
model with a linear component $\theta'Z$ and a nonparametric
component $g(V)$, and is identifiable because of the condition
(3.1). From this structure, we can see that when $V$ does not
contain the instrumental variable $W$ and $\alpha=\gamma$, the model
goes back to the original model (2.4) as $\xi $ is zero and $g(V)$
becomes the error term $\eta$ (if $\varepsilon$ is ignored). This
structure motivates our method. By introducing an instrumental
variable  $V$ so that $\xi$ has a zero conditional mean, and then we can
estimate $g(\cdot)$ to correct the bias occurred in the original
model. Although a nonparametric function $g(v)$ is involved, it will
be verified that the dimension $d+1$ of the variable $v$ is low.
Note that  for $V$, the key is to properly select $\alpha$  and $W$. From the
above description,  we can see that although $\alpha=\gamma$ should be a
natural and good choice, it is  unknown and
when the dimension is large, is cannot be estimated
consistently. Taking this into account, we first consider a general
$\alpha$ and construct a bias-corrected model with suitable $W$, or equivalently
a suitable matrix $A$.

Denote by $l=p-q$ the dimension of $\gamma$ and let
$\lambda=(0,\gamma_{2}-\frac{\gamma_1}{\alpha_1}\alpha_{2},\cdots,
\gamma_{l}- \frac{\gamma_1}{\alpha_1}\alpha_{l})'/\rho,$ where
$\alpha_1,\cdots,\alpha_l$ are the components of $\alpha$ and
$\alpha_1$ is supposed to be nonzero. We can ensure that, when $Z_i$
and $U_i$ satisfy
$$\lambda'E(U_i|Z_i,W_i)=\lambda'E(U_i|W_i),\eqno(3.3)$$
the model (3.2) is unbiased, i.e.,
$$E(\xi(V_i)|Z_i,V_i)=0.\eqno(3.4)$$ The proof of (3.4) will be presented in the Appendix.

When $(Z,U)$ is elliptically symmetrically distributed, the
condition (3.3) can be rewritten at population level as the following form:
$$\begin{array}{ll}\lambda'\Sigma_{U,Z^\star}A'(A\Sigma_{Z^\star,Z^\star}A')^{-1}
A(Z^\star-E(Z^\star))\vspace{1ex}\\=
\lambda'\Sigma_{U,Z^\star}B'(B\Sigma_{Z^\star,Z^\star}B')^{-1}B(Z^\star-E(Z^\star)),\end{array}\eqno(3.5)$$
where $\Sigma_{Z^\star,Z^\star}=Cov(Z^\star,Z^\star)$,
$\Sigma_{U,Z^\star} =Cov(U,Z^\star)$ and
$$B=\left(\begin{array}{llll}
I & 0&\cdots &0\\ A_1&a_{q+1}&\cdots& a_{q+d}\end{array}\right),$$
$A=(A_1,a_{q+1},\cdots,a_{q+d})$, $A_1$ is a $d\times q$ matrix and
$a_j,j=q+1,\cdots,a_{q+d},$ are $d$-dimensional column vectors.
Further, the  ellipticity condition can
be slightly weakened to be the following linearity condition:
$$E(U|C'Z^\star)=E(U)+
\Sigma_{U,Z^\star}C(C'\Sigma_{Z^\star,Z^\star}C)^{-1}C'(Z^\star-E(Z^\star))$$ for some given matrix $C$.
This linearity condition also results in (3.5). The linearity
condition has been widely assumed in the circumstance of high-dimensional
models. Hall and Li (1993) showed that it
often holds approximately when the dimension $p$ is high.

Under either the equation (3.3) or (3.5), the bias-corrected model
(3.2) is unbiased. Thus, we are now in the position to determine the
matrix $A$ by solving either the equation (3.3) or (3.5). A solution
is not difficult to be obtained. For example, if
$\Sigma_{Z^\star,Z^\star}=I_{q+d}$ and $B^{-1}$ exists, then we
choose $A$ satisfying
$$\Sigma_{U,Z^\star}A'(AA')^{-1}A(Z^\star-E(Z^\star))=
\Sigma_{U,Z^\star}(Z^\star-E(Z^\star)).\eqno(3.6)$$ It is known
that, if we can choose variables $U^{(1)},\cdots,U^{(d)}$ such that
the rank of matrix $\Sigma_{U,Z^\star}$ is $d$, then
$$\Sigma_{U,Z^\star}^+\Sigma_{U,Z^\star}=Q\left(\begin{array}{ll}I_d&0\\0&0\end{array}\right)
Q'=Q_1Q_1',$$ where $\Sigma_{U,Z^\star}^+$ is the Moore-Penrose
generalized inverse matrix of $\Sigma_{U,Z^\star}$, $I_d$ is a
$d\times d$ identify matrix, $Q=(Q_1,Q_2)$ an orthogonal matrix
satisfying $Q'Q=I_{q+d}$ and $Q_1'Q_1=I_d$. In this case, we choose
$$A=Q_1'.\eqno(3.7)$$ Such a matrix $A$ is a solution of (3.6) and thus a
solution of (3.5). With such a choice of $A$, the model (3.2) is
always unbiased whether the model (2.1) is sparse or not, the dimension of $\beta$
is high or low, and the choice of $\lambda_p$ is proper or not.

However, sometimes the matrix
$\Sigma_{U,Z^\star}^+\Sigma_{U,Z^\star}$ is unknown. Under this
situation, we will present a detailed procedure in Section 4 to
calculate $\Sigma_{U,Z^\star}^+\Sigma_{U,Z^\star}$ and $A$.
From the above choice of $A$, we can see that $g(v)$ is a
$d+1$-dimensional nonparametric function. If $d$ is large, we choose
a row vector to replace $A$ and will give a method in Section 4 to
find an approximate solution. With which, $g(v)$ is a 2-dimensional
nonparametric function.

The above deduction shows that the above bias-correction procedure
is free of the choice of $\alpha$. However, choosing a proper
$\alpha$ is of importance. It is clear that, combining (3.2) and
(3.3), choosing an $\alpha$ as close to $\gamma$ as possible should
be a good way although optimal choice leaves an unsolved and
interesting problem. In the estimation procedure, a natural choice
is  the value $\tilde \gamma^D$ for $\gamma$, which is obtained in
the Dantzig selection step. The details are presented in
Subsection~3.2 below. We will also discuss the asymptotic properties
of an estimation when we use a given $\alpha$ in the next
subsection.

\noindent{\bf 3.2 Asymptotic normality of estimation.} Throughout this
subsection we assume that the matrix $A$ satisfying (3.5) or (3.6)
has been obtained. Although the obtained $A$ is sometimes an
estimator rather than an exact solution, in this section we still
regard it as a nonrandom solution of (3.5) or (3.6) because such an
estimator is $\sqrt{n}$-consistent (see Section 4 below) and, as a
result, when $A$ is thought of a random vector, the theoretical
conclusions given below still hold.

Recall that the bias-corrected model (3.2) can be thought of as a partially
linear model. We therefore design an estimation procedure as
follows. First of all, as mentioned above, for any $\alpha$, the
model (3.2) is unbiased. Then we can design the estimation procedure
after $\alpha$ is determined by any empirical method. An empirical
choice $\alpha$ is designed as the Dantzig selector $\tilde\gamma^D$
of $\gamma$ determined by (2.2). Generally, given $\theta$ and for
any $\alpha$, the nonparametric function $g(v)$ is estimated by
$$g_\theta(v)=\frac{\sum_{k=1}^n(Y_k-\theta'
Z_k)L_H(V_k-v)} {\sum_{k=1}^nL_H(V_k-v)},$$ where $L_H(\cdot)$ is a
$(d+1)$-dimensional kernel function. Then $g_\theta(v)$ is a
$(d+1)$-variate nonparametric estimator. As was shown above, the
dimension $d+1$ is low. A simple choice of $L_H(\cdot)$ is a product
kernel as
$$L_H(V-v)=\frac{1}{h^{d+1}}K\Big(\frac{V^{(1)}-v^{(1)}}{h}\Big)\cdots
K\Big(\frac{V^{(d+1)}-v^{(d+1)}}{h}\Big),$$ where
$V^{(j)},j=1,\cdots,d+1$, are the components of $V$, $K(\cdot)$ is
an 1-dimensional kernel function and $h$ is the bandwidth depending
on $n$. Particularly, when $\alpha$ is chosen as $\tilde \gamma^D$,
we get an  estimator of $g(v)$ as
$$\hat g_\theta(v)=\frac{\sum_{k=1}^n(Y_k-\theta' Z_k)L_H(\Hat V_k-v)}
{\sum_{k=1}^nL_H(\Hat V_k-v)}$$ where $\hat
V=(U'\tilde\gamma^D/\hat\rho,W')'$ and
$\hat\rho=O(\|\tilde\gamma^D\|_{{\ell}_2}\sqrt{\lambda_M})$.

With these two estimations of $g(v)$, the
bias-corrected model (3.2) can be approximately expressed by the
following two models:
$$Y_i\approx \theta'Z_i+g_\theta(V_i)+\xi(V_i)\ \ \mbox{ and }\ \ Y_i\approx \theta'Z_i+\hat g_\theta(\Hat V_i)+\xi(\Hat V_i),$$ equivalently,
$$\tilde Y_i\approx \theta'\tilde Z_i+\xi(V_i)\ \ \mbox{ and }\ \ \hat Y_i\approx \theta'\hat Z_i+\xi(\Hat V_i),\eqno(3.8)$$
where
$$\tilde Y_i=Y_i-\frac{\sum_{k=1}^nY_kL_H(V_k- V_i)}
{\sum_{k=1}^nL_H( V_k-V_i)},\ \ \tilde
Z_i=Z_i-\frac{\sum_{k=1}^nZ_kL_H(V_k-V_i)}{\sum_{k=1}^nL_H( V_k-
V_i)},$$
$$\hat Y_i=Y_i-\frac{\sum_{k=1}^nY_kL_H(\Hat V_k-\Hat V_i)}
{\sum_{k=1}^nL_H(\Hat V_k-\Hat V_i)},\ \ \hat
Z_i=Z_i-\frac{\sum_{k=1}^nZ_kL_H(\Hat V_k-\Hat
V_i)}{\sum_{k=1}^nL_H(\Hat V_k-\Hat V_i)}.$$ Thus,  the sub-models
in (3.8) result in the estimations for $\theta$ as
$$\tilde\theta=S_n^{-1}\frac{1}{n}\sum_{i=1}^n
\tilde Z_i\tilde Y_i \ \ \mbox{ and }\ \
\hat\theta=S_n^{-1}\frac{1}{n}\sum_{i=1}^n \hat Z_i\hat
Y_i,\eqno(3.9)$$ where $S_n=\frac{1}{n}\sum_{i=1}^n\tilde Z_i\tilde
Z_i'$ or $S_n=\frac{1}{n}\sum_{i=1}^n\hat Z_i\hat Z_i'$,
respectively. Here we assume that the bias-corrected model (3.2) is
homoscedastic, that is $Var(\xi(V_i))=\sigma_V^2$ or $Var(\xi(\hat
V_i))=\sigma_V^2$ for all $i=1,\cdots,n$. If the model is
heteroscedastic, we respectively modify the above estimators as
$$\tilde\theta^*={S_n^*}^{-1}\frac{1}{n}\sum_{i=1}^n
\frac{1}{\sigma_i^2(V_i)}\tilde Z_i\tilde Y_i
 \ \ \mbox{ and }\ \ \hat\theta^*={S_n^*}^{-1}\frac{1}{n}\sum_{i=1}^n
\frac{1}{\sigma_i^2(\hat V_i)}\hat Z_i\hat Y_i,$$ where
$S_n^*=\frac{1}{n}\sum_{i=1}^n\frac{1}{\sigma_i^2(V_i)}\tilde
Z_i\tilde Z_i'$ or
$S_n^*=\frac{1}{n}\sum_{i=1}^n\frac{1}{\sigma_i^2(\hat V_i)}\hat
Z_i\hat Z_i'$, respectively, and $\sigma_i^2(V_i)=Var(\xi( V_i))$
and $\sigma_i^2(\hat V_i)=Var(\xi(\Hat V_i))$. Here
$\sigma_i^2(V_i)$ and $\sigma_i^2(\hat V_i)$ are supposed to be
known. If they are unknown, we can use their consistent estimators
to replace them; for details about how to estimate them see for
example H\"ardle {\it et al.} (2000). In the following we only
consider the estimators defined in (3.9). Finally, the estimators of
$g(v)$ can be defined as either $g_{\tilde\theta}(v)$ or $\hat
g_{\hat\theta}(v)$.

To study the consistency of the estimations, the following conditions for the model (3.2)  are assumed:

\begin{itemize}
\item[(C1)] The first two derivatives of $g(v)$ and $\xi(v)$ are continuous.
\item[(C2)] Kernel function
$K(\cdot)$ satisfies $$\int K(u)du=1, \int
u^jK(u)du=0,j=1,\cdots,k-1,0< \int u^kK(u)du<\infty.$$
\item[(C3)] $nh^{2(d+1)}\rightarrow\infty$.
\end{itemize}

Obviously, the conditions (C1)-(C3) are commonly used for
semiparametric models.  Under these conditions, the following
theorem provides the consistency of the bias-corrected estimator
$\tilde\theta$.

{\bf Theorem 3.1} {\it Assume that the conditions (C1)-(C3) hold,
and for given $\alpha$, (3.1) and (3.3) are satisfied. When $q$ is
fixed, and $p$ may be larger than $n$, then, as
$n\rightarrow\infty$,
$$\sqrt{n}(\tilde\theta-\theta)\stackrel{D}\longrightarrow N(0,\sigma_V^2S^{-1}),$$
where $S=E\{(Z-E(Z|V))(Z-E(Z|V))'\}$.}

{\bf Remark 3.1} For simplicity of presentation, in this theorem we only give the
the asymptotic normality for the case with fixed $q$. In fact, when
$q$ tends to infinity at a certain rate, the asymptotic normality still holds for every component
of $\theta$ (see for example Lam and Fan, 2008). This is because, after
bias-correction, the model (3.2) is indeed a partially linear model and then the proof can be similar with more technical and tedious details.
The proof of this theorem is postponed to the Appendix. The results in the theorem
show that the new estimator $\tilde\theta$ is $\sqrt{n}$-consistent
regardless of the choice of the shrinkage tuning parameter
$\lambda_p$ and thus it is convenient to be used in
practice. 
Furthermore, by the theorem and the commonly used nonparametric
techniques, we can prove that $g_{\tilde\theta}(v)$ is also
consistent. In effect, we can obtain the strong consistency and the
consistency of the mean squared error under some stronger
conditions. The details are omitted in this paper.


To investigate the asymptotic properties for the second estimator
$\hat\theta$ in (3.9) that is based on the Dantzig selector
$\tilde\gamma^D$, we need the following more conditions:

\begin{itemize}
\item[(C4)] The bandwidth $h$ is optimally chosen, i.e., $h=O(n^{-1/(2(k+d+1))})$.
\item[(C5)] Suppose that there exists a vector, say $\alpha$, such that
$\|\alpha\|_{{\ell}_2}\geq c$ for a positive constant $c$ and
$\|\tilde\gamma^D-\alpha\|_{{\ell}_2}/\|\tilde\gamma^D\|_{{\ell}_2}=O_p(n^{-\mu})$
for some $\mu$ satisfying $$1/2-k/(2(k+d+1))\leq\mu\leq 1/2.$$
\end{itemize}

As was stated in the previous sections,   $\alpha$ was an
arbitrary vector. The vector $\alpha$ in the condition (C5)
is then different. But for the simplicity of
representation we still use the same notation $\alpha$ in different
appearance. The condition (C5) is the key for the following
theorem and corollary. This condition does not mean that the Dantzig
selector $\tilde\gamma^D$ is consistent. Note that
$\|\tilde\gamma^D\|_{{\ell}_2}$ is large in
non-sparse case, and the accuracy of the solution of linear programm
can guarantee that $\|\tilde\gamma^D-\alpha\|_{{\ell}_2}$ is
relatively small  for the true value
of linear programm (2.2) at population level (see for example Malgouyres and Zeng, 2009). These
show that the condition (C5) is reasonable. 
   Both (C4) and (C5) can actually be weakened, but for the simplicity of technical proof
and presentation, we still use the current conditions in this paper.

{\bf Theorem 3.2} {\it Under the conditions (C1)-(C5), (3.1) and
(3.3), we have the following asymptotic representation for the
second estimator in (3.9):
$$\sqrt{n}(\hat\theta-\theta)=S^{-1}\frac{1}{\sqrt{n}}\sum_{i=1}^n\Big(
\tilde Z_i\tilde g( V_i)+\tilde Z_i\tilde \xi( V_i)\Big)+o_p(1),$$
where $S=E\{(Z-E(Z|V))(Z-E(Z|V))'\}$ and
$$\begin{array}{ll}\tilde g(V_i)=g(V_i)-\frac{\sum\limits_{k=1}^ng( V_k)L_H(V_k-V_i)}
{\sum\limits_{k=1}^nL_H(V_k-V_i)},\vspace{1ex}\\  \tilde \xi(
V_i)=\xi(V_i)-\frac{\sum\limits_{k=1}^n\xi(V_k)L_H( V_k- V_i)}
{\sum\limits_{k=1}^nL_H(V_k-V_i)},\vspace{1ex}\\ \tilde
Z_i=Z_i-\frac{\sum\limits_{k=1}^nZ_kL_H(V_k-
V_i)}{\sum\limits_{k=1}^nL_H(V_k- V_i)}.\end{array}$$ }

The proof of the theorem is given in the Appendix. From Theorem~3.2,
and Theorem 2.1.2 of H\"ardle {\it et al} (2000), the asymptotic
normality follows directly. The following corollary states the
detail.

{\bf Corollary~3.3} {\it Under the conditions of Theorem~3.1, when $q$
is fixed but $p$ may be larger than $n$, then, as
$n\rightarrow\infty$,
$$\sqrt{n}(\hat\theta-\theta)\stackrel{D}\longrightarrow N(0,\sigma_V^2S^{-1}).$$
}

As aforementioned in Remark 3.1, for the sub-model with diverging
$q$, the asymptotic normality can still hold under some stronger
conditions, the details are omitted here.


\noindent{\bf 3.3 Prediction.}
Together the estimation consistency with the adjusted sub-model (3.2),  we obtain an improved
prediction as
$$\hat
Y=\hat\theta'Z+\hat g_{\hat\theta}(V)\eqno(3.10)$$ and the
corresponding prediction error is
$$\begin{array}{lll}E(Y-\hat Y)^2&=&E((\hat\theta-\theta)'Z)^2+E(\hat
g_{\hat\theta}(V)-g(V))^2+E(\xi^2(
V))\vspace{1ex}\\&&+2E((\hat\theta-\theta)'Z(\hat
g_{\hat\theta}(V)-g(V)))+2E((\hat\theta-\theta)'Z\xi(
V))\vspace{1ex}\\&&+2E((\hat g_{\hat\theta}(V)-g(V))\xi(
V))\vspace{1ex}\\&=&E(\xi^2( V))+o(1).\end{array}$$
Such a prediction is of a smaller prediction error than the one by the classical Dantzig selector, and interestingly it is no need
with any high-dimensional nonparametric estimation.

In contrast, if we use the new estimator $\hat\theta$ and the sub-model (2.4),
rather than the adjusted sub-model (3.2), to construct prediction,
the resulting  prediction is defined as
$$\hat Y_S=\hat\theta'Z+\bar{\hat g}_{\hat\theta},\eqno(3.11)$$
where $$\bar{\hat g}_{\hat\theta}=\frac{1}{n}\sum_{i=1}^n\hat
g_{\hat\theta}(V_i).$$ For prediction, we need to add $\bar{\hat
g}_{\hat\theta}$ in (3.11) because the sub-model
(2.4) has a bias $E(g(V))$, otherwise, the prediction error would be even larger.  In this case, $\bar{\hat g}_{\hat\theta}$ is free of the predictor $U$ and the
resultant prediction (3.11) only uses the predictor $Z$ in the
sub-model (2.4). This  is different from the prediction
(3.10) that depends on both the low-dimensional predictor $Z$ and
high-dimensional predictor $U$. Thus (3.11)
is a sub-model based prediction. 
The
corresponding prediction error is
$$\begin{array}{lll}E(Y-\hat
Y_S)^2&=&E((\hat\theta-\theta)'Z)^2+E(\bar{\hat
g}_{\hat\theta}-g(V))^2+E(\xi^2(
V))\vspace{1ex}\\&&+2E((\hat\theta-\theta)'Z(\bar{\hat
g}_{\hat\theta}-g(V)))+2E((\hat\theta-\theta)'Z\xi(
V))\vspace{1ex}\\&&+2E((\bar{\hat g}_{\hat\theta}-g(V))\xi(
V))\vspace{1ex}\\&=&E(\xi^2( V))+Var(g(V))+2E(E(g(V))-g(V))\xi(
V))+o(1).\end{array}$$
This error is usually  larger than that of the prediction (3.10). But,
$$|E(E(g(V))-g(V))\xi(
V))|\leq (Var(g(V))Var(\xi(V)))^{1/2}$$ and usually the values of
both $Var(g(V))$ and $Var(\xi(V))$ are small. Then such a prediction
still has a smaller prediction error than the one obtained by the
sub-model (2.4) and the common LS estimator $\tilde\theta_S=({\bf
Z}'{\bf Z})^{-1}{\bf Z}'{\bf Y}$  as: $$\tilde
Y_S=\tilde\theta_S'Z\eqno(3.12)$$ with the corresponding error as
$$\begin{array}{lll}E(Y-\tilde Y_S)^2&=&E((\tilde\theta_S-\theta)'Z)^2+E(\gamma'U)^2+\sigma^2
+2E((\tilde\theta_S-\theta)'Z\gamma'U).\end{array}$$ Because
$\tilde\theta_S$ does not tend to $\theta$, the values of both
$E((\tilde\theta_S-\theta)'Z)^2$ and
$2E((\tilde\theta_S-\theta)'Z\gamma'U)$ are large and as a result
the prediction error is large.

The above results show that in the scope of prediction, the new
estimator can reduce prediction error under both the adjusted
sub-model (3.2) and the original sub-model (2.4). We will see that
the simulation results in Section~5 coincide with these conclusions.

\

\noindent {\large\bf 4. Calculation for $A$}

\noindent{\bf 4.1 Calculation of $A$ for the case with unknown
$\Sigma_{U,Z^\star}^+\Sigma_{U,Z^\star}$}. In the previous section, we suggested a simple choice of $A$ for the
case with known $\Sigma_{U,Z^\star}^+\Sigma_{U,Z^\star}$. We now
introduce an approach for choosing vector $A$ such that (3.6) holds
for the case with unknown $\Sigma_{U,Z^\star}^+\Sigma_{U,Z^\star}$.
For the convenience of representation, we here suppose $E(Z)=0$,
E(U)=0 and $Cov(Z^\star)=I$. In this case, (3.6) can be rewritten as
$$\Sigma_{U,Z^\star}A'(AA')^{-1}AZ^\star=\Sigma_{U,Z^\star}Z^\star.\eqno(4.1)$$

We denote
$\Sigma_{U,Z^\star}'\Sigma_{U,Z^\star}=\Omega=(\omega_{ij})$ with
$$\begin{array}{ll}\omega_{ij}=\sum_{k=1}^lE(U^{(k)}Z^{(i)})E(U^{(k)}Z^{(j)}),
 i,j\leq q,\vspace{1ex}\\
\omega_{i,q+s}=\omega_{q+s,i}=\sum_{k=1}^lE(U^{(k)}Z^{(i)})E(U^{(k)}U^{(s)}),
i=1,\cdots,q,s=1,\cdots,d,\vspace{1ex}\\
\omega_{q+r,q+s}=\sum_{k=1}^lE(U^{(k)}U^{(r)})E(U^{(k)}U^{(s)}),r,s=1,\cdots,d,\end{array}$$
where $Z^{(i)}$ and $U^{(k)}$ are the components of $Z$ and $U$,
respectively. It is known that $\Omega$ can be decomposed as
$$\Omega=Q\,\mbox{diag}\{\phi_1,\cdots,\phi_{d},0,\cdots,0\} Q',$$ where
$\phi_k,k=1,\cdots,d,$ are the positive eigenvalues of $\Omega$ and
$Q$ is the orthogonal matrix. Note that $l$ depends on $n$ and tends
to infinity as $n\rightarrow\infty$. To get  consistent estimator of
$Q$,  we need the following condition
$$\#\left\{\begin{array}{ll}E(U^{(i)}Z^{(j)}),E(U^{(k)}U^{(s)}):
E(U^{(i)}Z^{(j)})\neq 0,E(U^{(k)}U^{(s)})\neq 0,\\\hspace{4.6cm}
\mbox{ for all } i,j,k,s
\end{array}\right\}\leq C \eqno(4.2)$$ for a positive constant $C$,
where $\#\{S\}$ denotes the number of elements in the set $S$. Also
we can use some weaker conditions to replace (4.2). In fact the
conditions we need are similar to those required for
high-dimensional linear models, for example, the weak and strong
irrepresentable conditions (Zhao and Yu 2006) and the uniform
uncertainty principle (Cand\'es and Tao 2007). Note that $\Omega$ is
a low-dimensional matrix. Then, under the condition (4.2), $\Omega$
can be $\sqrt{n}$-consistently estimated; for example, a naive
estimator of $\hat \omega_{ij}$ for $i,j\leq q$ can be chosen as
$$\begin{array}{ll}\hat
\omega_{ij}\\=\sum\limits_{k=1}^l\frac{1}{n}\sum\limits_{s=1}^nU_s^{(k)}Z_s^{(i)}{\bf
1}\Big\{\frac{1}{n}\Big|\sum\limits_{s=1}^nU_s^{(k)}Z_s^{(i)}\Big|>
\frac{1}{\sqrt{n}}\Big\}\frac{1}{n}\sum\limits_{s=1}^nU_s^{(k)}Z_s^{(j)}{\bf
1}\Big\{\frac{1}{n}\Big|\sum\limits_{s=1}^nU_s^{(k)}Z_s^{(j)}\Big|>
\frac{1}{\sqrt{n}}\Big\},\end{array}$$ where ${\bf 1}\{S\}$ is the
indicator function of the set $S$. As was shown above, we can express
$\hat \Omega$ as
$$\hat\Omega=\hat Q\,\mbox{diag}\{\hat\phi_1,\cdots,\hat\phi_{d},0\cdots,0\}\hat
Q'\eqno(4.3)$$ and $\hat Q$ as $\hat Q=(\hat Q_1,\hat Q_2)$.
Finally, the estimator of $A$ is obtained by
$$\hat A=\hat Q_1.$$


\noindent{\bf 4.1 Calculation of $A$ for large $d$}. As we mentioned
before, when $d$ is large, the solution $A$ of (4.1) has $d$ columns
and then $(d+1)$-dimensional nonparametric estimation will be
involved, which leads an inefficient estimation. Thus, we consider
an approximation solution of (4.1), which is a row vector. Without
confusion, we still use the notation $A$ to denote this row vector.
That is, we choose a row vector $A$ such that
$$A'AZ^\star=\Sigma_{U,Z^\star}^+\Sigma_{U,Z^\star}Z^\star.\eqno(4.4)$$
The approximation solution is identical to the solution of (4.1) in
form as when $A$ is a row vector, recalling that it is normalized to
be norm one, $AA'=1$. In this case, to get a low-dimensional
nonparametric function $g(v)$, we choose $d=1$, i.e., $Z^*$ is a
$q+1$-dimensional vector. Similar to the above determination, when
$A$ is unknown, we can also construct an estimation as follows.
Denote $A=(a_1,\cdots,a_q,a_{q+1})$, $A_k=a_kA$ and
$\Sigma_{U,Z^\star}^+\Sigma_{U,Z^\star}=(D_1',\cdots,
D_{q}',D_{q+1}')'$, where $D_k,k=1,\cdots,q+1$, are
$(q+1)$-dimensional row vectors. Then we estimate $A$ via solving
the following optimization problem:
$$\inf\Big\{Q(a_1,\cdots,a_{q+1}):
\sum_{k=1}^{q+1}a_k^2=1\Big\},\eqno(4.5)$$ where
$Q(a_1,\cdots,a_{q+1})=\frac{1}{n}\sum_{i=1}^n\sum_{k=1}^{q+1}\|(A_k-D_k)Z_i^\star\|^2$.
By the Lagrange multiplier, we obtain the estimators of
$A_k,k=1,\cdots,q+1,$ as
$$\hat A_k=\Big(D_k\frac{1}{n}\sum_{i=1}^nZ_i^\star {Z_i^\star}'+cc_ke_k/2\Big)\Big(\frac{1}{n}
\sum_{i=1}^nZ_i^\star {Z_i^\star}'+c_k I\Big)^{-1},\eqno(4.6)$$
where $c_k>0$, which is similar to a ridge parameter, depends on
$n$ and tends to zero as $n\rightarrow\infty$, and $e_k$ is a row
vector with $k$-th component 1 and the others being zero. Note that
the constraint $\|A\|=1$ implies $\|A_k\|=\pm a_k$. Finally, by
combining (4.6) and this constraint we get an estimator of $a_k$ as
$$\hat a_k=\pm \|\hat A_k\|$$ and consequently the estimator of $A$
is obtained by
$$\hat A=(\hat a_1,\cdots,\hat a_q,\hat a_{q+1}).$$

\

\noindent {\large\bf 5. Simulation studies}

In this section we examine the performance of the new method by
simulations. By mean squared error (MSE), model prediction error
(PE) and their $std$\,MSE and $std$\,PE as well, we compare the method with
the Gaussian-dantzig selector first. In ultra-high dimensional
scenarios, the Dantzig selector cannot work well, we
use the sure independent screening (SIS) (Fan and Lv 2008) to bring
dimension down to a moderate size and then to make comparison with
the Gaussian-dantzig selector.  As is well known, there are several
factors that are of great impact on the performance of variable
selection methods: dimensions $p$ of covariate $X$,  correlation
structure between the components of covariate $X$, and variation of
the error which can be measured by theoretical model R-square
defined by $R^2=(Var(Y)-\sigma^2_\varepsilon)/Var(Y)$. In order to
comprehensively illustrate the theoretical conclusions and
performance, we design three experiments. The main goal of the first
experiment is to examine the effect of $R^2$ as the smaller $R^2$
is, the more difficult correctly selecting variables is. The second
experiment is to investigate the impact from the correlation between
the components of covariate $X$, and the third is to check whether the
two-step procedure of the SIS and the Dantzig selector works or not.

{\bf Experiment 1.} This experiment is designed mainly for: (1)
comparing the new estimator $\hat\theta$ defined by (3.9) with the
Gaussian-dantzig selector $\tilde\theta_{S}$; (2) examining the
effect of different choices of the theoretical model $R^2$ of the
full model (2.1); (3) checking the effect of the correlation between
the components of $X$ when $R^2$ is fixed. To achieve these goals,
we compare the MSEs, the PEs and their $std$\,MSE and $std$\,PE of
the two different estimators $\hat\theta$ and $\tilde\theta_{S}$,
and the two models (2.4) and (3.2). In the simulation,  to determine
the regression coefficients in our simulation, we decompose the
coefficient vector $\beta$ as two parts: $\beta_I$ and $\beta_{-I},$
where $I$ denotes the set of locations of significant components of
$\beta_I$, and let $S=|I|$ denote the number of elements contained
in $I$. Three types of $\beta_I$ are considered:
\\ Type (I): $\beta_I= (1,0.4,0.3,0.5,0.3,0.3,0.3)'$
and $I$= \{1,2,3,4,5,6,7\};\\ Type (II): $\beta_I=
(1,0.4,0.3,0.5,0.3,0.3,0.3)'$ and $I= \{1,17,33,49,65,81,97\}$;\\
Type (III): $\beta_I= (1,0.4,-0.3,-0.5,0.3,0.3,-0.3)'$ and $I=
\{1,2,3,4,5,6,7\}$. \\
As it is very rare that all other coefficients are
exactly zero, non-sparse models are considered. To mimic practical scenarios, we set the values of the
components $\beta_{-Ii}$'s of $\beta_{-I}$ as follows. Before
performing the variable selection and estimation, we generate
$\beta_{-Ii}$'s from uniform distribution $\mathcal {U}(-0.5,0.15)$
and the negative values of them are then set to be zero. After the
coefficient vector $\beta$ is determined, we consider it as a fixed
value vector and regard $\beta_I$ as the main part of the
coefficient vector $\beta$. We use this way to set the values of
$\beta_{-Ii}$'s because in the simulations below, there are too many
insignificant variables with small/zero coefficients and it makes
little sense to give a common value for them. As too  many values
for these insignificant coefficients, we do not list all of them
here. We use $\hat I$ to denote the set of subscript of coefficients
$\theta$ in $\beta$, that is the coefficients' subscript of
variables selected into sub-model. we  assume $X \thicksim
N_{p}(\mu,\Sigma_{X})$, with $\mu$  the components corresponding to
$I$  are 0 and others are 2   and   the $(i,j)$-th element
$\Sigma_{ij}=(-\rho)^{\mid i-j\mid}$, $0<\rho<1.$ Furthermore, the
error term $\varepsilon$ is assumed to be normally distributed as
$\varepsilon \thicksim N(0,\sigma^2)$. In this experiment, we choose
different $\sigma$ to obtain different type of full model with
different $R^2$. In the simulation procedure and the kernel function
is chosen to be Gaussian kernel
$K(u)=\frac{1}{\sqrt{2\pi}}\exp\{-\frac{u^2}{2}\}$. In this
experiment, the choice of parameter $\lambda_p$ in the Dantzig
selector is just like that given by Cand\'es and Tao (2007), which
is the empirical maximum of $|X'z|_i$ over several realizations of
$z\sim N(0,I_n).$

The following Tables 1 and 2 report the MSEs and the corresponding
PEs via 200 repetitions. In these tables,  $\hat Y$ is the
prediction via the adjusted model (3.2) that is based on  the full
dataset, $\hat Y_S$ is the prediction via the sub-model (2.4) with
the new estimator $\hat\theta$ defined in (3.9), $\tilde Y_S$ stands
for the prediction via the sub-model (2.4) and the Gaussian-dantzig
selector $\tilde\theta_{S}$. For the definitions of $\hat Y$, $\hat
Y_S$ and $\tilde Y_S$ see (3.10), (3.11) and (3.12), respectively.
The purpose of such a comparison is to see whether the adjustment
works and whether we should use the sub-model (2.4) when the
high-dimensional data are not available (say, too expensive to
collect), whether the new estimator $\hat\theta$ together with the
sub-model (2.4) is helpful for prediction accuracy. The sample size
is $50$, and for the prediction, we perform the experiment with 200
repetitions to compute the proportion $\tau$ of which the prediction
error of $\hat Y_S$ is less than that of $\tilde Y_S$ in the 200
repetitions. The larger $\tau $ is, the better the new estimator is.
We have the following considerations in designing the experiment:
a). We will study models with the theoretical model $R^2$ ranging
between 0.3 and 1.0, which can be determined by the value of the
variance of error term $\sigma^2$, here we choose $\sigma^2$=0.2,
0.6, 0.9, 1.3 and 1.9 respectively; b). The correlation between the
components of $X$ should have effect for the estimation, we then
consider different correlation coefficients $0.1$ and $0.7$.

1. Let $n=50,p=100,S=7$ and  $\rho=0.1$ .  For each type of $\beta$,
we choose different $\sigma$ to
control the theoretical $R^2$ and consider  five cases.\\
For type (I), we have the following results:\\ {\small
Case 1.  $R^2=0.98$, $I=\{1,2,3,4,5,6,7\}$ and $\hat I=\{1 ,2,3,4, 6, 7 \}$;\\
Case 2.  $R^2=0.82$, $I=\{1,2,3,4,5,6,7\}$ and $\hat I=\{1  ,2,4  ,6 ,7,55\}$;\\
Case 3.  $R^2=0.67$, $I=\{1,2,3,4,5,6,7\}$ and $\hat I=\{ 1,  2, 3,  4, 15, 22, 28,81
\}$;\\
Case 4.  $R^2=0.50$, $I=\{1,2,3,4,5,6,7\}$ and $\hat I=\{ 1,2, 4,  27, 29, 49, 53,  84 \}$;\\
Case 5.  $R^2=0.31$, $I=\{1,2,3,4,5,6,7\}$ and $\hat I=\{1,4 , 5 ,24 ,25 ,42 ,43  ,  62 \}$.
}\\
For type (II), we have the following results:\\ {\small
Case 1.  $R^2=0.98$, $I=\{1,17,33 ,49, 65,81, 97\}$ and $\hat I=\{1,17, 33,49,65, 81, 97 \}$;\\
Case 2.  $R^2=0.84$, $I=\{1,17,33 ,49, 65,81, 97\}$ and $\hat I=\{1,17, 33, 43, 49,  81
\}$;\\
Case 3.  $R^2=0.71$, $I=\{1,17,33 ,49, 65,81, 97\}$ and $\hat I=\{ 1,15, 17, 33, 49, 62, 72 \}$;\\
Case 4.  $R^2=0.53$, $I=\{1,17,33 ,49, 65,81, 97\}$ and $\hat I=\{1, 5,26,29,33,43,49,53,65,74       \}$;\\
Case 5.  $R^2=0.35$, $I=\{1,17,33 ,49, 65,81, 97\}$ and $\hat I=\{ 1,7,17,26,29,31,49,72,80,96, 97,98
\}$.
}\\
For type (III), we have the following results:\\ {\small
Case 1.  $R^2=0.98$, $I=\{1,2,3,4,5,6,7\}$ and $\hat I=\{ 1  ,   2  ,   3   ,  4 ,    5 ,    6       \}$;\\
Case 2.  $R^2=0.83$, $I=\{1,2,3,4,5,6,7\}$ and $\hat I=\{ 1, 3 , 4,  5,   6, 7 , 15
\}$;\\
Case 3.  $R^2=0.69$, $I=\{1,2,3,4,5,6,7\}$ and $\hat I=\{ 1, 2,4, 5,7, 92
\}$;\\
Case 4.  $R^2=0.51$, $I=\{1,2,3,4,5,6,7\}$ and $\hat I=\{1,5, 7, 8,  67, 71
\}$;\\
Case 5.  $R^2=0.33$, $I=\{1,2,3,4,5,6,7\}$ and $\hat I=\{1,4, 6, 7, 21, 23, 38, 50, 75,  83 \}$.
}\\

\begin{center}
{ \small \centerline{{\bf Table 1.}     MSE, PE and their standard
errors with $n=50,p=100,S=7$ and $\rho=0.1$} \tabcolsep0.04in
\vspace{-1ex}
\newsavebox{\tablebox}
\begin{lrbox}{\tablebox}
\begin{tabular}{cc|cc|ccc|c}
  \hline\hline
  & &\multicolumn{2}{c|}  {MSE($std$\,MSE)}  &\multicolumn{3}{c|}{PE($std$\,PE)}& \\
 type&$R^2$&$\hat \theta$   &$\tilde\theta_S$  &$\hat Y$& $\hat Y_S$ & $\tilde Y_S$ &\raisebox{1.5ex}[0pt]{$\tau$}\\\hline
&0.98 &0.0032(0.0118) &0.0866(0.3519) &0.1630(0.0405) &0.2299(0.0535) &1.1587(0.5549) &    200/200 \\
&0.82 & 0.0134(0.0544)&0.1197(0.1654) &0.6603(0.1497) &0.7249(0.1564) &1.4755(0.3475) &    200/200 \\
(I)&0.67 &0.0273(0.1288) &     0.0430(0.1283) & 1.3038(0.2952) &1.3438(0.3018) &1.4821(0.3266) &166/200 \\
&0.50 &     0.0543(0.2387) &0.0694(0.2221) &2.5371(0.5500) &2.5919(0.5633) &2.7176(0.6020) &        142/200 \\
&0.31 &0.1028(0.4689) & 0.1131(0.4876) &4.9199(1.1856) & 4.9960(1.2070) &5.0708(1.1965) &        126/200 \\
\hline
&0.98 &0.0052(0.0202) &0.3540(1.4263) & 0.2584(0.0569) &0.2744(0.0583) &1.1324(2.4262) &        200/200 \\
&0.84 & 0.0162(0.0686) & 0.4087(0.3730) & 0.8310(0.1823) &0.8417(0.1834) &3.7996(0.7909) &        200/200 \\
(II)&0.70 & 0.0292(0.1112) &0.1770 (0.2559) & 1.4761(0.3028) &1.4727(0.3018) & 2.6389(0.5804) &        199/200 \\
&0.53 & 0.0588(0.3024) & 0.0942(0.2988) & 2.8825(0.6534) & 2.8700(0.6460) &3.2707(0.6758) &        171/200 \\
&0.35 & 0.1107(0.6896) & 0.1251(0.6368) & 5.4055 (1.1809) &5.3896(1.1856) &5.6004(1.2280) &        141/200 \\
\hline
&0.98 &0.0028(0.0113) &  0.0879(0.2938) &0.1643(0.0410) &0.2365(0.0537) &1.2282(0.5590) &        200/200 \\
&0.83 & 0.0114(0.0531) &0.0873(0.1589) &0.5874 (0.1332) & 0.6938(0.1533) &1.3483(0.3118) &        200/200 \\
(III)&0.69 & 0.0234(0.0934) &0.1294(0.1667) &1.1922(0.2857) & 1.2445(0.2961) &1.9950(0.4379) &        196/200 \\
&0.51 & 0.0529(0.1715) & 0.0913(0.1775) &  2.6373(0.5788) & 2.7418(0.6098) & 2.9601(0.6288) &        164/200 \\
&0.33 &0.1006(0.5013) & 0.1083(0.5158) & 5.0952(1.2099) &5.1720(1.2241) &5.2372(1.2594) &        119/200 \\
\hline\hline
\end{tabular}
\end{lrbox}
\scalebox{0.9}{\usebox{\tablebox}} }
\end{center}

The simulation results are reported in Table 1. The results suggest
that  the adjustment of
(3.2) works very well, the corresponding estimation and prediction
are uniformly the best  among the  competitors. Further, as we
mentioned, when the full dataset is not available and we thus use
the sub-model of (2.4), the new estimator $\hat \theta$ is also
useful for prediction. It can be seen that $\hat Y_S$ is better than
$\tilde Y_S$, and the value of $\tau$ is
larger than 0.7 in 13 cases out of 15 cases and in
the other 2 cases, it is larger than or about 0.6. 

 2. To provide more information, we also consider the case with
higher correlation $\rho=0.7$: $n=50,p=100,S=7$. Also
 different $\sigma$'s are chosen to
control the theoretical $R^2$.\\
For type (I), we consider the following five cases.\\ {\small
Case 1.  $R^2=0.96$, $I=\{1,2,3,4,5,6,7\}$ and $\hat I=\{1,2,4,5,6,7\}$;\\
Case 2.  $R^2=0.71$, $I=\{1,2,3,4,5,6,7\}$ and $\hat I=\{ 1 , 2 , 4 , 81\}$;\\
Case 3.  $R^2=0.53$, $I=\{1,2,3,4,5,6,7\}$ and $\hat I=\{ 1,4,8,9\}$;\\
Case 4.  $R^2=0.35$, $I=\{1,2,3,4,5,6,7\}$ and $\hat I=\{1 ,4 , 8,51 \}$;\\
Case 5.  $R^2=0.20$, $I=\{1,2,3,4,5,6,7\}$ and $\hat I=\{ 1, 2 , 6
,84\}$.
}\\
For type (II), we consider the following five cases.\\ {\small
Case 1.  $R^2=0.98$, $I=\{1,17,33 ,49, 65,81, 97\}$ and $\hat I=\{1,17,33 ,49, 65, 97\}$;\\
Case 2.  $R^2=0.84$, $I=\{1,17,33 ,49, 65,81, 97\}$ and $\hat I=\{1,18 ,49 ,65 ,97\}$;\\
Case 3.  $R^2=0.69$, $I=\{1,17,33 ,49, 65,81, 97\}$ and $\hat I=\{ 1, 2 ,49,52 ,65\}$;\\
Case 4.  $R^2=0.52$, $I=\{1,17,33 ,49, 65,81, 97\}$ and $\hat I=\{1,15, 33 , 49,  76 ,84 ,98 \}$;\\
Case 5.  $R^2=0.34$, $I=\{1,17,33 ,49, 65,81, 97\}$ and $\hat
I=\{1,2 ,24 , 48 , 49,55,87 , 97\}$.
}\\
For type (III), we consider the following five cases.\\ {\small
Case 1.  $R^2=0.96$, $I=\{1,2,3,4,5,6,7\}$ and $\hat I=\{ 1,2 ,4 ,6,7\}$;\\
Case 2.  $R^2=0.74$, $I=\{1,2,3,4,5,6,7\}$ and $\hat I=\{ 1,4 ,6 ,7\}$;\\
Case 3.  $R^2=0.56$, $I=\{1,2,3,4,5,6,7\}$ and $\hat I=\{1,6 , 7 ,33,56\}$;\\
Case 4.  $R^2=0.38$, $I=\{1,2,3,4,5,6,7\}$ and $\hat I=\{1,4 ,7 ,51,93\}$;\\
Case 5.  $R^2=0.23$, $I=\{1,2,3,4,5,6,7\}$ and $\hat I=\{ 1 ,2,7
,31  ,  45  ,  80  ,  85,    88\}$. }

\newpage

\begin{center}
{ \small \centerline{{\bf Table 2.}     MSE, PE and their standard
errors with $n=50,p=100,S=7$ and $\rho=0.7$} \tabcolsep0.04in
\vspace{-1ex}
\begin{lrbox}{\tablebox}
\begin{tabular}{cc|cc|ccc|c}
  \hline\hline
  & &\multicolumn{2}{c|}  {MSE($std$\,MSE)}  &\multicolumn{3}{c|}{PE($std$\,PE)}& \\
 type&$R^2$&$\hat \theta$   &$\tilde\theta_S$  &$\hat Y$& $\hat Y_S$ & $\tilde Y_S$ &\raisebox{1.5ex}[0pt]{$\tau$}\\\hline
&0.96 &0.0136(0.0504) &    0.3285(0.4226) &    0.2472(0.0517) &    0.2706(0.0599) &    1.7397(0.3804) &        200/200 \\
& 0.71 &    0.0253(0.1426) &    0.0709(0.2401) &    0.6530(0.1463) &    0.6945(0.1557) &    1.9892(0.2070) &        197/200 \\
(I)&       0.53 &    0.0373(0.1621) &    0.1108(0.2310) &    1.2779(0.2744) &    1.3235 (0.2861) &    1.5985(0.3736) & 177/200 \\
&       0.35 &    0.0613(0.3122) &    0.0999(0.3289) &    2.3431(0.5342) &    2.3694(0.5395) &    2.6339(0.5799) &        161/200 \\

           &        0.2 &    0.1198(0.6479) &    0.1292(0.6619) &    5.1184(1.2643) &    5.1347(1.2729) &    5.1764(1.2420) &        129/200 \\\hline

           &       0.98 &    0.0122(0.0484) &     0.2730(0.3789) &     0.2648(0.0730) &     0.2809(0.0757) &     1.1952(0.2440)& 200/200 \\

           &       0.84 &    0.0201(0.0924) &    0.1799(0.2037) &    0.6567(0.1453) &    0.6580(0.1452) &    1.6477(0.3560) & 200/200 \\

           (II)&       0.69 & 0.0303(0.1338) &    0.2899(0.4442) & 1.2955(0.2992) &    1.2996(0.3047) &    2.7125(0.5861) & 200/200 \\

           &       0.52 &    0.0644(0.3395) &    0.1141l(0.4388) &    2.5572(0.5558) &    2.5633(0.5582) &    3.2790(0.6834) & 191/200 \\

           &       0.34 &    0.1245(0.5615) &    0.1831(0.6787) &    5.0731(1.1850) &    5.0818(1.1743) &    5.5988(1.2782) & 161/200 \\\hline

           &       0.96 &    0.0239(0.0626) &0.6020(2.1653) &    0.2596(0.0560) &    0.2897(0.0630) &   1.6754(1.4970) &        200/200 \\

           &       0.74 &    0.0315(0.1158) &    0.4401(0.5248) &    0.6435(0.1435) &    0.6485(0.1442) &    2.7859(0.6035) &  200/200 \\

           (III)&       0.56 &    0.0749(0.2373) &    0.1736(0.2679) &    1.3334(0.2947) &    1.4367(0.3217) &    1.8643(0.3965) & 189/200 \\

           &       0.38 &    0.0687(0.3227) &    0.1701(0.3809) &    2.3637(0.4538) &    2.4645(0.4818) &    2.9415(0.5992) & 178/200 \\

           &       0.23 &    0.1740(0.8078) & 0.2446(0.8718) &    4.8488(1.1812) &    4.8887(1.1968) &    5.1471(1.1499) &   145/200 \\

\hline\hline
\end{tabular}
\end{lrbox}
\scalebox{0.9}{\usebox{\tablebox}} }
\end{center}

\

Table~2  shows that when $\rho$ is larger, the conclusions about the
comparison are almost identical to those presented in Table~1; 
Thus it concludes that no matter $\rho$ is larger or not, our new
method always works quite well.

We are now in the position to make another comparison. In Experiments~2
and 3 below, we do not use the data-driven approach as given in
Experiment 1 to select $\lambda_p$, while manually select several
values to see whether our method works or not. This is because in
the two experiments, it is not our goal to study shrinkage tuning
parameter, but is our goal to see whether the new method works after
we have a sub-model.

{\bf Experiment 2.}  In this experiment, our focus is   how the
correlation between variables affects the estimations. The
distribution of $X$ is the same as that in Experiment~1 except for
the dimension. 
The coefficient vector $\beta$ is designed as type (I) in Experiment
1.
 Furthermore, the error term
$\varepsilon$ is assumed to be normally distributed as $\varepsilon
\thicksim N(0,0.2^2)$.  

As different choices of $\lambda_p$ will usually lead to different
sub-models, equivalently,  to different  estimators $\hat I$ of $I$,
we are then able to examine, when the numbers of significant
variables that are included into the submodels are different, the
performance of the new estimation by MSE and PE. 
In this experiment,
we consider two cases with two values of $\lambda$. 
The setting is as follows. For $n=50,p=100,S=7$,
$\rho=0.1,0.3,0.5,0.7.$ We consider two cases
for each $\rho$: \\
$\rho=0.1:\\
$ Case 1. $\lambda_p=3.97,$ $I$=\{1,2,3,4,5,6,7\}, $\hat I$=\{1,     2,     3,     4,     5,     6,     7 \}\\
Case 2. $\lambda_p=6.53,$ $I$=\{1,2,3,4,5,6,7\}, $\hat I$=\{1,     3,     4,     6,    95 \}\\
$\rho=0.3:\\$ Case 1. $\lambda_p=3.32,$ $I$=\{1,2,3,4,5,6,7\}, $\hat I$=\{1,     2,     3,     4,     5,     6 \}\\
Case 2. $\lambda_p=6.77,$ $I$=\{1,2,3,4,5,6,7\}, $\hat I$=\{ 1,     2,     4,     6,    23 \}\\
$\rho=0.5:\\$ Case 1. $\lambda_p=3.72,$ $I$=\{1,2,3,4,5,6,7\}, $\hat I$=\{1,     2,     4,     5,     6,     7 \}\\
Case 2. $\lambda_p=7.29,$ $I$=\{1,2,3,4,5,6,7\}, $\hat I$=\{1,     4,     5,     7,    41,    58,    72 \}\\
$\rho=0.7:\\$ Case 1. $\lambda_p=3.50,$ $I$=\{1,2,3,4,5,6,7\}, $\hat I$=\{1,     3,     4,     7,    41,    75\}\\
Case 2. $\lambda_p=7.22,$ $I$=\{1,2,3,4,5,6,7\}, $\hat I$=\{1,     4,     7,    51,    64,    67,    68,    83 \}\\

\newpage
\begin{center}
{ \small \centerline{{\bf Table 3.}     MSE, PE and their standard
errors with $n=50,p=100,S=7$ } \tabcolsep0.045in \vspace{-1ex}
\begin{lrbox}{\tablebox}
\begin{tabular}{cc|cc|ccc|c}
  \hline\hline
  & &\multicolumn{2}{c|}  {MSE($std$\,MSE)}  &\multicolumn{3}{c|}{PE($std$\,PE)}& \\
$\rho$&Case&$\hat \theta$   &$\tilde\theta_S$  &$\hat Y$& $\hat Y_S$&$\tilde Y_S$ &\raisebox{1.5ex}[0pt]{$\tau$}\\\hline
                          &1&   0.0052(0.0242) &    0.2929(0.3877) &   0.2580(0.0528) &   0.2612(0.0527) &   3.0195(0.6691) &  200/200\\
\raisebox{1.5ex}[0pt]{0.1}&2 &   0.0104(0.0357) &   0.2347(0.1784) &   0.5135(0.1074) &   0.6430(0.1282) &    5.921(0.4172) &200 /200 \\\hline
                          &1 &   0.0070(0.0289) &   0.4067(1.6692) &   0.2732(0.0590) &   0.3324(0.0735) &   5.6406(1.8289) &200/200\\
\raisebox{1.5ex}[0pt]{0.3}&2 &     0.0163(0.0458) &   0.5048(0.4107) &   0.4048(0.0881) &   0.5014(0.1078) &   6.4471(0.7697) &200/200\\\hline
                          &1 &   0.0079(0.0336) &   0.4826(1.9425) &    0.2436(0.0551) &    0.3053(0.0674) &   5.8204(1.8152) & 200/200\\
\raisebox{1.5ex}[0pt]{0.5}&2 &   0.0136(0.0512) &    0.1532(0.1835) &    0.3655(0.0841) &    0.4245(0.0914) &    6.4357(0.3262) &200/200\\\hline
                          &1 &   0.0157(0.0602) &    0.2296(0.2970) &    0.2688(0.0580) &    0.3198(0.0711) &    6.6313(0.3560) &200/200\\
\raisebox{1.5ex}[0pt]{0.7}&2 &   0.0149(0.0637) &    0.1914(0.1420) &    0.2974(0.0624) &    0.3225(0.0672) &    7.5435(0.1169) &197/200\\
\hline\hline
 \end{tabular}

\end{lrbox}
 \scalebox{0.9}{\usebox{\tablebox}} }
\end{center}

From Table~3, we can see clearly that the correlation is of impact
on the performance of the variable selection methods:  the
estimation gets worse with larger $\rho$. However, the new method
uniformly works much better than the Gaussian Dantzig selector, when
we compare the performance of the methods with different values of
$\lambda$ and then with different sub-models. We can see that in
case I, the sub-models are more accurate than those in case II in
the sense that they can contain more significant variables we want
to select. Then, the estimation based on the Gaussian Dantzig
selector can work better and so can the new method. Note that $\rho$ is about 1 meaning that in all the 200 repetitions, $\hat Y \le \hat Y_S.$

In the following, we consider ultra high-dimensional data.

{\bf Experiment 3.} For very large $p$, the Dantzig selector method
alone cannot work well. Thus, we use the sure independent screening
(SIS,  Fan and Lv 2008) to reduce the number of variables to a
moderate scale that is below the sample size, and then perform the
variable selection and parameter estimation afterwards by the
Gaussian Dantzig selector and our adjustment method. 
We first consider  $n=100,p=1000$  and $S=10$ with $\rho$=0.1, 0.5
and 0.9 respectively, and for each $\rho$  two
$\lambda_p$ are used to obtain two $\hat I$ as follows. \\
{\small
 For $\rho$=0.1,
$\beta_I=(1.0, -1.5, 2.0, 1.1, -3.0, 1.2, 1.8, -2.5, -2.0, 1.0)'$, consider two cases:  \\
Case 1. $\lambda_p$=4.50,  $I$=\{1,2,3,4,5,6,7,8,9,10\}, $\hat I=\{1,     3  ,   5 ,    6 ,    7  ,   8,     9 ,  318,   514 ,  723,   760
\}$;\\
Case 2. $\lambda_p$=7.30,  $I$=\{1,2,3,4,5,6,7,8,9,10\}, $\hat I=\{ 2 ,    3 ,    5,     8,   515 ,  886
 \}$.\\
For $\rho$=0.5, $\beta_I=(1.0, -1.5, 2.0, 1.1, -3.0, 1.2, 1.8, -2.5, -2.0, 1.0)'$, consider two cases:\\
Case 1. $\lambda_p$=3.56,  $I$=\{1,2,3,4,5,6,7,8,9,10\}, $\hat I=\{1  ,   2  ,   5 ,    7,     8 ,    9 ,  846 ,  878 ,  976
  \}$;\\
Case 2.$\lambda_p$=6.92,  $I$=\{1,2,3,4,5,6,7,8,9,10\}, $\hat I=\{ 2   ,  3 ,    5 ,    8 ,   10,   882 ,  963
 \}$.\\
For $\rho$=0.9, $\beta_I=(1.0, -1.5, 2.0, 1.1, -3.0, 1.2, 1.8, -2.5, -2.0, 1.0)'$, consider two cases:\\
Case 1. $\lambda_p$=1.80,  $I$=\{1,2,3,4,5,6,7,8,9,10\}, $\hat I=\{  3,     5,     8,    10 ,  415 ,  432
 \}$;\\
Case 2.$\lambda_p$=5.83,  $I$=\{1,2,3,4,5,6,7,8,9,10\}, $\hat I=\{ 2
,   3   ,  5 ,  114 ,  121 ,  839 ,  853,   882 ,  984\}$.}

With this design, the $\lambda$ in case 1 results in that  more
significant variables are selected into the sub-model than those in
case 2 so that we can see the performance of the adjustment method.

\

\begin{center}
{ \small \centerline{{\bf Table 4.}     MSE, PE and their standard
errors with $n=100,p=1000,S=10$ } \tabcolsep0.04in \vspace{-1ex}
\begin{lrbox}{\tablebox}
\begin{tabular}{cc|cc|ccc|c}
  \hline\hline
  & &\multicolumn{2}{c|}  {MSE($std$\,MSE)}  &\multicolumn{3}{c|}{PE($std$\,PE)}& \\
$\rho$&Case&$\hat \theta$   &$\tilde\theta_S$  &$\hat Y$& $\hat
Y_S$&$\tilde Y_S$ &\raisebox{1.5ex}[0pt]{$\tau$}\\\hline
                          &1&  0.7588(0.3497) &    71.4031(7.5501) &     6.8104(1.5485) &     8.0107(1.6574) &    94.7515(19.2968) & 200/200 \\
\raisebox{1.5ex}[0pt]{0.1}&2&  0.8523(0.5343) &   122.8426(15.0952)
&    13.1274(2.7772) &    16.0812(3.4160) &   189.7134(34.8081) &
200/200 \\\hline
                          &1&  3.6170(1.1823) &   104.8420(13.5089) &     9.9151(1.9902) &    11.2352(2.2316) &   133.4762(26.5058) &200/200\\
\raisebox{1.5ex}[0pt]{0.5}&2&3.4771(1.2683) &    92.3485(12.5122) &
11.6643(2.6704) &    12.7811(2.8941) &   134.3821(24.4896)
&200/200\\\hline
                          &1&  5.9027(2.7039) &   107.6118(23.4383) &     8.2842(1.6181) &    11.3518(2.1745) &   148.3143(27.4828) & 200/200\\
\raisebox{1.5ex}[0pt]{0.9}&2& 3.8963(2.1760) &    59.1525(11.3152) &    10.8033(2.1411) &    12.9395(2.4835) & 68.7272(13.4061) & 200/200\\
\hline\hline
 \end{tabular}

\end{lrbox}
\scalebox{0.85}{\usebox{\tablebox}} }
\end{center}

\

From Table 4, we can see that the SIS does work to reduce the
dimension so that the Gaussian Dantzig selector and our method can
be performed. Whether the correlation coefficient is small or large
(the values of $\rho$ change from 0.1 to 0.9), the new method works
better than the Gaussian Dantzig selector. The conclusions are
almost identical to those when $p$ is much smaller in Experiments 1
and 2. Thus, we do not give more comments here. Further, when
comparing
 the results of case 1 and case 2, we can see that the adjustment can work better
 when the submodel is not well selected. The value of $\rho=1.$

 In the following we  check the effect of model
size when the dimension is larger. In doing so, we choose $n=150,
p=2000$, $\rho=0.3$ with $S=5, 10$. For each $S$ we choose two
$\lambda_p$ to obtain two $\hat I.$ \\
For $S$=5,
$\beta_I=(4.0, -1.5, 6.0, -2.1, -3.0)'$, we consider two cases:  \\
Case 1. $\lambda_p$=3.45, $I$=\{1,2,3,4,5\}, $\hat I=\{ 1,2,3,4,5,15,1099,1733
 \}$;\\
Case 2. $\lambda_p$=8.36, $I$=\{1,2,3,4,5\}, $\hat I=\{1,3,554,908
  \}$.\\
For $S$=10, $\beta_I=(4.0, -1.5, 6.0, -2.1, -3.0, 1.2, 3.8, -2.5, -2.0, 7.0)'$, consider two cases:\\
Case 1. $\lambda_p$=3.02, $I$=\{1,2,3,4,5,6,7,8,9,10\}, $\hat I=\{1,2,3,5,7,8,9,10,1701
 \}$;\\
Case 2. $\lambda_p$=9.08, $I$=\{1,2,3,4,5,6,7,8,9,10\}, $\hat I=\{1,3,5,7,8   \}$.\\

\begin{center}
{ \small \centerline{{\bf Table 5.}     MSE, PE and their standard
errors with $n=150,p=2000,\rho=0.3$ } \tabcolsep0.02in \vspace{-1ex}
\begin{lrbox}{\tablebox}
\begin{tabular}{cc|cc|ccc|c}
  \hline\hline
  & &\multicolumn{2}{c|}  {MSE($std$\,MSE)}  &\multicolumn{3}{c|}{PE($std$\,PE)}& \\
$$S$$&Case&$\hat \theta$   &$\tilde\theta_S$  &$\hat Y$& $\hat
Y_S$&$\tilde Y_S$ &\raisebox{1.5ex}[0pt]{$\tau$}\\\hline
                          &1&  0.4245(0.2102) &262.6392(21.2109) & 6.4015(1.3038) &     6.3439(1.2879) &322.9945(62.6228) &200/200 \\
\raisebox{1.5ex}[0pt]{5}&2&  1.9510(1.0923) & 359.5838(32.4150) & 24.1959(4.8932) &    24.8013(5.1629) & 559.3584(98.1216) &200/200 \\\hline
                          &1&  0.8799(0.5108) & 498.7862(59.0383) &10.6009(2.3903) &    12.3505(2.6381) &946.3400(175.1009) & 200/200\\
\raisebox{1.5ex}[0pt]{10}&2&  1.8524(0.7599) &68.1862(43.3612) &15.0471(2.8069) & 16.9161 (3.1755) &  1623.4936(111.5972) & 200/200\\
\hline\hline
 \end{tabular}

\end{lrbox}
\scalebox{0.9}{\usebox{\tablebox}} }
\end{center}

The results in Table 5 show that the SIS is again useful for reducing the
dimension for the use of the Gaussian Dantzig selector and our
method. When the model size is smaller, estimation accuracy can be
better with smaller MSE and PE. In other words, when the model size
is smaller, variable selection can perform better and sub-model can
be more accurate (case 1 with $S=5$), the adjustment method does not
have much help, and in contrast, it is useful for improving
estimation accuracy when the
sub-model is very different from the full model. 

In summary, the results in the five tables above clearly show the
superiority of the new estimator $\hat\theta$ and the new sub-model
(3.2)/the sub-model (2.4) over the others in the sense with smaller
MSEs, PEs and standard errors, and large proportion $\tau$. The good
performance holds for different combinations of the sizes of
selected sub-models (values of $\lambda_p$), $n,p,S,I$, $R^2$ and
the correlation between the components of $X$. The new method is
particularly useful when a submodel, as a working model, is very
different from underlying true model. Thus,  the adjustment method
is very worth of recommendation. However, as a trade-off, the
adjustment method involves nonparametric estimation, although
low-dimensional ones, it might not be that helpful when the sub-model
is accurate enough. Thus, we may consider using it after a check
whether the submodel is significantly biased. The relevant research
is ongoing.

\

\noindent {\large\bf Appendix}

{\it Proof of (3.4)} Note that
$$\begin{array}{ll}&E(\xi(V)|Z,V)\vspace{1ex}
\\&=E(Y-\theta'Z-g(V)|Z,V)\vspace{1ex}
\\&=E(Y-\theta'Z|Z,V)-E(g(V)|Z,V)
\vspace{1ex}\\&=E(\gamma'U+\varepsilon|Z,V)-E(E(\gamma'U+\varepsilon|V)|Z,V)
\vspace{1ex}\\&=\gamma'E(U|Z,V)-\gamma'E(U|V)\vspace{1ex}\\&=\gamma'E(U|Z,\alpha'U/\rho,W)
-\gamma'E(U|\alpha'U/\rho,W).\end{array}$$ Further,
$$\begin{array}{ll}E(\gamma'U/\rho|Z=z,\alpha'U/\rho=t,W=w)\vspace{1ex}\\
=E(\gamma'U/\rho|Z=z,U^{(1)}=(t\rho-\sum_{j=2}^{l}U^{(j)}\alpha_{j})/\alpha_1,W=w)
\vspace{1ex}\\
=E(\frac{\gamma_1}{\alpha_1\rho}(t\rho-\sum_{j=2}^{l}U^{(j)}\alpha_{j})
+\sum_{j=2}^{l}U^{(j)}\gamma_{j}/\rho|Z=z,W=w)\vspace{1ex}\\
=E(\frac{\gamma_1}{\alpha_1}t
+\sum_{j=2}^{l}U^{(j)}(\gamma_{j}-\frac{\gamma_1}{\alpha_1}\alpha_{j})/\rho|Z=z,W=w)\vspace{1ex}\\
=\frac{\gamma_1}{\alpha_1}t+\lambda'E(U|Z=z,W=w).
\end{array}$$
Similarly,
$E(\gamma'U/\rho|W=w,U'\gamma/\rho=t)=\frac{\gamma_1}{\alpha_1}t+\lambda'E(U|W=w)$.
Combining the above results leads to
$$E(\xi(V)|Z,V)=\rho\lambda'(E(U|Z,W)-E(U|W)),$$ as required. \hfill $\fbox{}$

\

{\it Proof of Theorem 3.1} \ The proof follows directly from the
unbiasedness of the model (3.2) for any $\alpha$ and Theorem 2.1.2
of H\"ardle {\it et al} (2000). \hfill $\fbox{}$

\

{\it Proof of Theorem 3.2} \ Denote $V^*=({\gamma^*}'U/\rho^*,W')'$
and $\rho^*=O(\|\gamma^*\|_{{\ell}_2}\sqrt{\lambda_M})$, where
$\gamma^*$ is a $l$-dimensional vector between $\tilde\gamma^D$ and
$\alpha$. Then there exists a vector $\gamma^*$ such that
$$\begin{array}{ll}\frac{1}{n}\sum\limits_{k=1}^nL_H(\Hat V_k-v)
=\frac{1}{n}\sum\limits_{k=1}^nL_H(V_k-v)+\frac{1}{n}\sum\limits_{k=1}^n\dot
L_H(V^*_k-v)(\Hat V_k-V_k),\end{array}$$ where $\dot L_H(\cdot)$ is
the derivative of $L_H(\cdot)$. By the conditions (C1) and (C5), we
have
$$\Hat V_k-V_k=O_p(n^{-\mu})$$ and, consequently,
$$\begin{array}{ll}\frac{1}{n}\sum\limits_{k=1}^n\dot
L_H(V^*_k-v)(\Hat V_k-V_k)=O_p(n^{-\mu}).\end{array}$$ By standard nonparametric technique, see
H\"ardle, et al (2000), it is easy to have
$$\begin{array}{ll}\frac{1}{n}\sum\limits_{k=1}^nL_H(V_k-v)-f_V(v)
=O_p\Big(h^k+\frac{1}{\sqrt{nh^{2(d+1)}}}\Big),\end{array}$$ and
then
$$\begin{array}{ll}\frac{1}{n}\sum\limits_{k=1}^nL_H(\hat V_k-v)
-f_V(v)=O_p\Big(h^k+\frac{1}{\sqrt{nh^{2(d+1)}}}\Big)+O_p(n^{-\mu}),
\end{array}$$ where $f_V$ is the density function of $V$.
Similarly, we can prove
$$\frac{1}{n}\sum\limits_{k=1}^nZ_kL_H(\Hat V_k-v)-
\int z
f_{Z,V}(z,v)dz=O_p\Big(h^k+\frac{1}{\sqrt{nh^{2(d+1)}}}\Big)+O_p(n^{-\mu}),$$
where $f_{Z,V}$ is the joint density function of $(Z,V)$. Combining
the results above leads to $\hat
Z=Z-E(Z|V)+O_p\Big(h^k+\frac{1}{\sqrt{nh^{2(d+1)}}}\Big)+O_p(n^{-\mu})$
and, consequently,
$$S_n-S=O_p\Big(h^k+\frac{1}{\sqrt{nh^{2(d+1)}}}\Big)+O_p(n^{-\mu}).$$ Further, by the definition of $\hat\theta$ and the
above result, we have
$$\hat\theta-\theta=S^{-1}\frac{1}{n}\sum_{i=1}^n\Big(
\hat Z_i\hat g(\Hat V_i)+\hat Z_i\hat \xi(\Hat
V_i)\Big)\Big\{1+O_p\Big(h^k+\frac{1}{\sqrt{nh^{2(d+1)}}}\Big)+O_p(n^{-\mu})\Big\},$$
where
$$\begin{array}{ll}\hat g(\Hat V_i)=g(\Hat V_i)-\frac{\frac{1}{n}\sum_{k=1}^ng(\Hat V_k)L_H(\Hat V_k-\Hat V_i)}
{\frac{1}{n}\sum_{k=1}^nL_H(\Hat V_k-\Hat V_i)},\vspace{1ex}\\  \hat
\xi(\Hat V_i)=\xi(\Hat V_i)-\frac{\frac{1}{n}\sum_{k=1}^n\xi(\Hat
V_k)L_H(\Hat V_k-\Hat V_i)} {\frac{1}{n}\sum_{k=1}^nL_H(\Hat
V_k-\Hat V_i)}.\end{array}$$ Again by the conditions (C1) and (C4),
we have
$$\begin{array}{ll}\hat g(\Hat V_i)=g(V_i)-\frac{\frac{1}{n}\sum_{k=1}^ng( V_k)L_H(V_k-V_i)}
{\frac{1}{n}\sum_{k=1}^nL_H(V_k-V_i)}+O_p(n^{-\mu})=\tilde g(V_i)+O_p(n^{-\mu}),\vspace{1ex}\\
\hat \xi(\Hat
V_i)=\xi(V_i)-\frac{\frac{1}{n}\sum_{k=1}^n\xi(V_k)L_H( V_k- V_i)}
{\frac{1}{n}\sum_{k=1}^nL_H(V_k-V_i)}+O_p(n^{-\mu})=\tilde\xi(V_i)+O_p(n^{-\mu}),\vspace{1ex}\\
\hat Z_i=Z_i-\frac{\frac{1}{n}\sum_{k=1}^nZ_kL_H( V_k-
V_i)}{\frac{1}{n}\sum_{k=1}^nL_H(V_k- V_i)}+O_p(n^{-\mu})=\tilde
Z_i+O_p(n^{-\mu}).\end{array}$$ Note that, under the condition (C4),
$$\begin{array}{ll}\tilde g(V_i)=O_p\Big(h^k+\frac{1}{\sqrt{nh^{2(d+1)}}}\Big)=O_p(n^{-k/(2(k+d+1))}),\vspace{1ex}\\
\tilde\xi(V_i)=O_p\Big(h^k+\frac{1}{\sqrt{nh^{2(d+1)}}}\Big)=O_p(n^{-k/(2(k+d+1))}),
\vspace{1ex}\\ \tilde
Z_i=O_p\Big(h^k+\frac{1}{\sqrt{nh^{2(d+1)}}}\Big)=O_p(n^{-k/(2(k+d+1))}).\end{array}$$
Therefore combining the above results can complete the proof of the
theorem. \hfill $\fbox{}$

{\it Proof of Corollary 3.3} \ The proof follows directly from the
result of Theorem 3.2 and Theorem 2.12 of H\"ardle {\it et al}
(2000). \hfill $\fbox{}$


\

\leftline{\large\bf References}

\begin{description}

\baselineskip=15pt

\item Bai, Z. and Saranadasa, H. (1996). Effect of high dimension:
by an example of a two sample problem. {\it Statistica Sinica}, {\bf
6}, 311-329.

\item Cand\'es, E. J. and Tao, T. (2005). Decoding by linear
programming. {\it IEEE Trans. Inform. Theory} {\bf 51}, 4203-4215.

\item Cand\'es, E. J. and Tao, T. (2006). Near-optimal signal recovery
from random projections: Universal encoding strategies? {\it IEEE
Trans. Inform. Theory} {\bf 52}, 5406-5425.

\item Cand\'es, E. J. and Tao, T. (2007). The Dantzig selector:
statistical estimation when $p$ is much larger than $n$. {\it Ann.
Statist.} {\bf 35}, 2313-2351.

\item Chen, S. X. and Qin, Y. L. (2010). A two sample test for high
dimensional data with applications to gene-set testing. {\it Ann.
Statist.}, {\bf 38}, 808-835.

\item Diaconis, P. and Freedman, D. (1984). Asymptotics of graphical
projection pursuit. {\it Ann. Statist.}, {\bf 12}, 793-815.

\item Donoho, D. L. (2000). High-dimensional data analysis: The
curses and blessing of dimensionality, Lecture on August 8, 2000, to
the American Mathematical Society on ``Math Challenges of the 21st
Century".

\item Fan, J. and Peng, H. (2004). Nonconcave penalized likelihood
with a diverging number of parameters. {\it Ann. Statist.}, {\bf
32}, 928-961.

\item Fan, J., Peng, H. and Huang, T. (2005). Semilinear
high-dimensional model for normalized of microarray data: a
theoretical analysis and partial consistency. {\it J. Am. Statist.
Ass.}, (with discussion), {\bf 100}, 781-813.

\item Fan, J. and Lv, J. (2008). Sure independence screening for
ultrahigh dimensional feature space. {\it J. R. Statist. Soc.} B
{\bf 70}, 849-911.

\item H\"ardle, W., Liang, H. and Gao, T. (2000). {\it Partially linear
models.} Physica Verlag.

\item
Hall, P. and Li, K. C. (1993). On almost linearity of low
dimensional projection from high dimensional data. {\it Ann.
Statist.} {\bf 21}, 867-889.


\item Huang, J., Horowitz, J. L. and Ma, S. (2008). Asymptotic
properties of Bridge estimators in sparse high-dimensional
regression models. {\it Ann. Statist.}, {\bf 36}, 2587-2613.

\item Huber, P. J. (1973). Robust regression: Asymptotic,
conjectures  and Montes Carlo. {\it Ann. Statist.}, {\bf 1},
799-821.

\item James, G. M. and Radchenko, P. (2009). A generalized Dantzig
selector with shrinkage tuning. {\it Biometrika}, {\bf 96}, 323-337.

\item James, G. M. , Radchenko, P. and Lv, J. C. (2009).
Dasso: connections between the Dantzig selector and lasso. {\it
Journal of the Royal Statistical Society, Series B}, {\bf 71},
127-142.

\item James, G. M. and Radchenko, P. (2009). A generalized Dantzig
selector with shrinkage tuning. {\it Biometrika} {\bf 96}, 323-337.

\item Kettenring, J., Lindsay, B. and Siegmund, D., eds. (2003).
Statistics: Challenges and opportunities for the twenty-first
century. NSF report. Available at
www.pnl.gov/scales/docs/nsf\_report.pdf

\item Kosorok, M. and Ma, S. (2007). Marginal asymptotics for the
large $p$, small $n$ paradigm: With application to Microarray data.
{\it Ann. statist.}, {\bf 35}, 1456-1486.

\item Kuelbs, J. and Anand, N. W. (2010). Asymptotic inference for
high dimensional data. {\it Ann. Statist.}, {\bf 38}, 836-869.

\item Lam, C and Fan, J. (2008). Profile-kernel likelihood inference
with diverging number of parameters. {\it Ann. Statist.}, {\bf 36},
2232-2260.

%

\item Li, G. R., Zhu, L. X. and Lin, L. (2008). Empirical Likelihood for a
varying coefficient partially linear model with diverging number of
parameters. {\it Manuscript}.

\item Malgouyres, F. and Zeng, T. (2009). A predual proximal point algorithm solving a non negative basis pursuit denoising model. {\it Int. J. comput Vis.}, {\bf 83}, 294-311.

%

\item Portnoy, S. (1988). Asymptotic behavior of likelihood methods
for exponential families when the number of parameters tends to
infinity. {\it Ann. Statist.}, {\bf 16}, 356-366.
%

\item
Robins, J. and Vaart, A. V. D. (2006). Adaptive nonparametric
confidence sets. {\it Ann. Statist.}, {\bf 34}, 229-253.

\item Van Der Lanin, M. J. and Bryan, J. (2001). Gene expression
analysis with parametric bootstrap. {\it Biostatistics}, {\bf 2},
445-461.


\item Zhao, P. and and Yu, B. (2006). On model selection consistency
of Lasso. {\it Journal of Machine Learning Research}, {\bf 7},
2541-2563.

\end{description}

\end{document}